\documentclass[a4paper,fleqn]{cas-sc}

\usepackage[authoryear,longnamesfirst]{natbib}

\def\tsc#1{\csdef{#1}{\textsc{\lowercase{#1}}\xspace}}
\tsc{WGM}
\tsc{QE}
\tsc{EP}
\tsc{PMS}
\tsc{BEC}
\tsc{DE}

\begin{document}
\let\WriteBookmarks\relax
\def\floatpagepagefraction{1}
\def\textpagefraction{.001}
\shorttitle{Comprehensive Text Mining on Various Sacred Scriptures}
\shortauthors{YOUNOUS MOFENJOU PEURIEKEU et~al.}

\title [mode = title]{A Text Mining Discovery of Similarities and Dissimilarities Among Sacred Scriptures}                  
\tnotemark[1,2]

\tnotetext[1]{This document is the results of the research
   project funded by the National Science Foundation.}

\tnotetext[2]{The second title footnote which is a longer text matter
   to fill through the whole text width and overflow into
   another line in the footnotes area of the first page.}

\author[1]{Younous Mofenjou Peuriekeu}[type=editor,
                        auid=000,bioid=1,
                        orcid=0000-0001-7511-2910]
\cormark[1]
\ead{cvr_1@tug.org.in}
\ead[url]{www.cvr.cc, cvr@sayahna.org}

\credit{Conceptualization of this study, Methodology, Software}

\address[1]{School of Mathematical Sciences, African Institute for Mathematical Sciences, Crystal Garden, Limbe Cameroon}

\author[1]{Victoire Djimna Noyum}[style=chinese]
\fnmark[1]

\author[1]{Cyrille Feudjio}[style=chinese]
\fnmark[2]

\author%
[3]
{Alkan  G\"{o}ktug.}
\ead{rishi@stmdocs.in}
\ead[URL]{www.stmdocs.in}

\address[3]{School of Mathematical Sciences, ETH Zurich, Rämistrasse 101, 8092 Zurich, Switzerland}

\author[2]{Ernest Fokou\'e}[%
   role=,
   prefix=,
   ]

\ead{cvr3@sayahna.org}
\ead[URL]{www.sayahna.org}

\credit{Data curation, Writing - Original draft preparation}

\address[2]{School of Mathematical Sciences, Rochester Institute of Technology, Rochester, NY 14623, USA}

\cortext[cor1]{Corresponding author}
\cortext[cor2]{Principal corresponding author}
\fntext[fn1]{This is the first author footnote. but is common to third
  author as well.}
\fntext[fn2]{Another author footnote, this is a very long footnote and
  it should be a really long footnote. But this footnote is not yet
  sufficiently long enough to make two lines of footnote text.}

\nonumnote{This note has no numbers. In this work we demonstrate $a_b$
  the formation Y\_1 of a new type of polariton on the interface
  between a cuprous oxide slab and a polystyrene micro-sphere placed
  on the slab.
  }

\begin{abstract}
	The careful examination of sacred texts gives valuable insights into human psychology, different ideas regarding the organization of societies as well as into terms like \textit{truth} and \textit{God}. To improve  and deepen our understanding of sacred texts, their comparison and their separation is crucial. For this purpose, we use of our data set has nine sacred scriptures. This work deals with separation of the Quran, the Asian scriptures Tao-Te-Ching, the Buddhism, the Yogasutras and the Upanishads as well as the four books from the Bible, namely the Book of Proverbs, the Book of Ecclesiastes, the Book of Ecclesiasticus and the Book of Wisdom. These scriptures are analyzed based on the natural language processing NLP creating the mathematical representation of the corpus in terms of frequencies called document term matrix (DTM). After this analysis, machine learning methods like supervised and unsupervised learning are applied to perform classification. Here we use the Multinomial Naive Bayes (MNB), the Super Vector Machine (SVM), the Random Forest (RF) and the K-nearest Neighbors (KNN). We obtain that among these methods MNB is able to predict the class of a sacred text with an accuracy about 85.84 \%.
\end{abstract}




\maketitle

\section{Introduction}
The progress in transportation and communication that has brought all the people of the world into one global village has also brought the religions of the world into close contact. To know what is unique or specific about a religion and how religion has gained importance in the life of humans, it is helpful to understand the structural patterns of these texts. Generally, when we talk about sacred or holy text, we refer to the religious context. Sacred Scripture are passages from the religious traditions. Often, these scriptural passages support a common theme. This method of organization allows each topic to be addressed with the resources of many different traditions, often providing a broader and deeper understanding of the topic than considering only the resources of a single tradition. Each religion has much value to contribute to humankind's understanding of truth, which transcends any particular expression. The proper description of religion is an area of study that has been tackled on different standpoints. Some of the common approaches to the study of religion are through history, anthropology, psychology and sociology.

\begin{itemize}
	\item The historians have been interested in religion as a social movement and has traced the development of various religions.
	\item The anthropologists have been interested in the genetic approach in the study of man, both in
	physiological and psychological aspects.
	\item The sociologist is essentially interested in the institutional and the ritualistic aspects of religion.  
\end{itemize}
The transcendentalist approach shows that religion concerns values and ideas. This approach is different from other disciplines of social sciences because, while the latter studies religion only as a structure, the former studies the religious values constituting the inner core of religion.

There are many different fields studying religions. These studies are always based on investigations on the scriptures. Hence, understanding a religion requires an analysis of its religious scripture.

\subsubsection{The sacred texts of our study}
\begin{itemize}
	\item Quran:
	
	It is the single-authored, central religious text of Islam and written in the eastern Arabian dialect of Classical Arabic. The Quran has 30 divisions or Juz with 114 chapters as Surah, according to the length of surahs, but not according to when they were revealed and not by subject matter. Each Surah is subdivided into verses or Ayat. The Quran contains about 6,236 verses with 77,477 words. The Quran is believed to be orally revealed by God through the archangel Gabriel to Muhammad, considered as the final prophet by Muslims. These three corpora are highly unstructured and do not follow or adhere to structured or regular syntax and patterns. Before applying any statistical technique or machine learning algorithm, the corpus needs to be converted into a structured text or a vector format that these techniques and algorithms can work with \citep{sah2019asian}.
	\item Asian text:
	\begin{itemize}
		\item Yogasutras: From India
		
		This Book contains essence of wisdom. It states that humans think of themselves as living a purely physical life in their material bodies. The central claim is that, in reality, they have gone far indeed from pure physical life; for ages, their life has been psychical, and they have been centered and immersed in the psychic nature \citep{sah2019asian}.
		
		\item Buddhism: 
		
		This book teaches the so-called \textit{four noble truth}. Each of these truths entails a duty: stress is to be comprehended, the origination of stress is to be abandoned, the cessation of stress is to be realized, and the path to the cessation of stress is aimed to be developed. When all of these duties have been fully performed, it is believed that the mind gains total release\citep{sah2019asian}.
		
		\item Tao Te Ching: Which is from China
		
		Tao Te Ching is a Chinese classic text traditionally credited to the $6^{th}$ century BC sage Laozi. The Tao Te Ching is a short text which has two parts, the Tao Ching (chapters 1-37) and the Te Ching ( chapters 38-81), which may have been edited together into the received text,
		possibly reversed from an original Te Tao Ching. The chapters talk about staying detached, letting go and keeping things simple.
		
		\item  The Upanishads:
		
		This book represents the loftiest heights of ancient Indo-Aryan thought and culture. They form the wisdom portion or
		Gnana-Kanda of the Vedas, as contrasted with the Karma-Kanda or sacrificial portion. From each of the four great Vedas known as Rik, Yajur, Sama and Atharva, there is a large portion which deals predominantly with rituals and ceremonials aiming to teach humans how they can prepare themselves for higher attainment  \citep{sah2019asian}.
	\end{itemize}

	\item Christianity books: Their origin is from Central Asia/America
	\begin{itemize}
		\item Book of Proverbs:
		
		The Book of Proverbs is the Book of where we have the wise sentences regulating the morals of men and directing them to  virtue.
		
		\item Book of Ecclesiasticus:
		
		From this Book of Ecclesiasticus, it brings out the remarkable lessons of the virtues.
		\item Book of Ecclesiastes :
		
		Also called the preacher, the Book of Ecclesiastes is one of the Christian Books, where Solomon plays a very important role as a preacher.
		
		\item Book of Wisdom:
		
		The Book of Wisdom abounds with instructions and exhortations to kings and all magistrates to minister justice in the commonwealth, teaching all kinds of virtues under the general names of justice and wisdom \citep{sah2019asian}.
	\end{itemize}
\end{itemize}

\subsection{Text analysis and learning}

To define text analysis, we can say that, it is the technique used to model and structure the information content of textual sources for investigation, exploratory data analysis, and research. This technique is described by the set of statistical and machine learning approach. This technique is very useful in text mining to simplify the data analysis, research, and investigation\citep{hobbs1982natural}. It is an instance of text analysis applications.

On the other hand, text analysis also refers to the set of processes analyzing a very large amount of unstructured text data to come up with many attributes as topics respectively keywords. Hence, Natural Language Processing (NLP), Natural Language Understanding (NLU) and Data Mining (DM) are overlapping areas which include techniques to process large corpora and extract useful information\citep{elton2019using}. As technique of extraction, we have text pre-processing, text normalization, text categorization, text clustering, text similarity etc. However, the corpus are the collection of sentences with phrases and words. For a given unstructured text data, the foundation of an analysis is the text pre-processing and text normalization converting the raw corpus to structured data whereas, text clustering and text similarity are processes to check the degree of closeness of two corpora and the process of grouping similar documents respectively.

\subsection{Objectives}
Text mining is an approach that combines several fields of research and also several software tools. According to \citep{ignatow2017introduction}, this is a technique that has begun to develop in social sciences such as anthropology, communication \citep{ezzeldin2020metaresearching}, economics \citep{levenberg2014predicting}, education \citep{evison2013corpus}, and psychology \citep{sklad2012effectiveness}. Social scientists have spent many decades studying transcribed interviews, newspaper articles, speeches, and other forms of textual data. After discovering the new, more sophisticated and rapid method of text mining, they have begun to adapt this approach to different forms of textual data analysis. Text mining also takes into account computer science.  So, in our study, we will not limit ourselves to the use of this new analysis technique, but also study the extension of natural language mining.

The reason why we have come to a multiplicity of religious teachings is due to the fact that their sacred texts are different from one religion to another. The fact that they are different can be explained by the lessons they teach their followers, their period of appearance, and also their geographical location. In this study, we will deepen the excavation of sacred texts by using text mining techniques.

Besides analyzing the differences between the books, it is also crucial to extract their similarities. This work attempts to find the similarity using text mining techniques.

Automated lexical analysis techniques are used nowadays for retrieval of useful information from large amounts of unstructured texts. Several studies have been done in order to study different religions. Frank Lloyd Sindler et al. \citep{sindler2011comparative} have written a thesis on a comparative study of Christian, Jewish, and Islamic theodicy. No automated technique has been used to analyze the lexical content of religious texts and only three religions have been considered in this study. Altogether, very little efforts have been made in the past to automatize analysis of important religious texts. 

Daniel McDonald et al \citep{mcdonald2014text} presented a method for automated extraction of topics from nine religious texts to form a self-organizing map to find relationships between these religious texts. The backdraw of this study is that only nine were taken into consideration leaving out important world religions.

Buddhism, Jainism , Sikhism .Qahl, Salha Hassan Muhammed et al. \citep{qahl2014automatic}  developed an automatic similarity detection engine using the Bible and the Quran as corpus to explore the performance of various feature extraction and machine learning techniques. Only two religious texts were taken into consideration and it did not give deeper insight into differences regardings the ideas and beliefs proposed by these two religions.  

The most recent work was done by Preeti and al. \citep{sah2019asian} where height sacred text were used, four from the Bible and four from Asian religions to present the statistical machine learning approach and to analyze many aspects of sacred texts from both the Asian and Biblical scriptures. Thus, in our study, we will extend the amount of religious books by adding the Islamic sacred scripture and then use a more sophisticated and rapid method of text mining, namely the natural language processing (NLP) to extract the high quality information from those books. Based on this, we will explore the performance of various feature extraction methods and machine learning techniques.

\section{Text analysis and learning}

After getting the different sacred texts that we will use for this study, we proceed with analyzing them in a structured manner. We begin with the Part-Of-Speech (POS) which is the process of marking up the word in a corpus (text) as corresponding to the special part of the speech based on both its definition and its context. For our analysis we will use the package \textit{nltk} in Python. After presenting the \textit{Information Retrieval}, we give an overview about Natural Language Processing. Then we explain the different steps of the pre-processing.

\subsection{Information Retrieval}

Before stating the pre-processing, we need to understand some concepts related to knowledge discovery. Information Retrieval (IR)  means finding documents that contain information about a certain term respectively keyword \citep{manning1999foundations}. For instance, Google does such kind of retrieval in their search engine. This search engine uses query based algorithm to track the trends and attain more significant results. However, this application has many more everyday uses such as email search or searching a file on a personal computer.

\subsection{Natural Language Processing}

\textbf{Natural language processing} (NLP) refers to the automatic processing and analysis of unstructured text data. It is important to underline that NLP performs different types of analysis such as Named Entity Recognition (NER) for abbreviation and their synonyms extraction to find the relationships among them \citep{laxman2013improved}. From this approach, we can identify all the instances of specified objects from a group of documents. Moreover, it allows the identification of relationship and other information to attain their key concept. In real world, a single entity has numerous terms like TV and Television.

\subsection{Pre-processing}

To give an overview of what we call \textbf{pre-processing}, we can say that it is the conversion of raw text data to a structured sequence of linguistic components in the form of a standard vector. Examples of pre-processing techniques are:

\begin{itemize}
	\item Tokenization ;
	
	Tokenization here means the process of splitting words, sentences or texts into a list of tokens. The components with special syntax and semantics are called "tokens". We distinguish two tokenization techniques, namely word tokenization and sentence tokenization. Word tokenization is the technique of splitting sentences into constituent words and sentences tokenization is the process of splitting the text corpus to meaningful sentences.
	
	\item Stemming ; 
	
	Stemming means the reduction of words to their roots so that, for instance, the different grammatical forms or declinations of verbs are identified and counted as the same word \citep{nisbet2009handbook}. So basically this step is an important approach for information retrieval and text analysis
	applications. An example is clustering and automatic text processing. 
	
	\item Lemmatization.
	
	Lemmatization is a process to resolve words to their dictionary form. In fact, a lemma of a word is its dictionary form. To resolve a word to its lemma, its part-of-speech is needed. Another aspect to note about lemmatization is that it is often times harder to create a lemmatizer in a new language than stemming.
\end{itemize}

\begin{figure}[htbp!] 
	\centering
	\includegraphics[height=12.5cm, width=13cm]{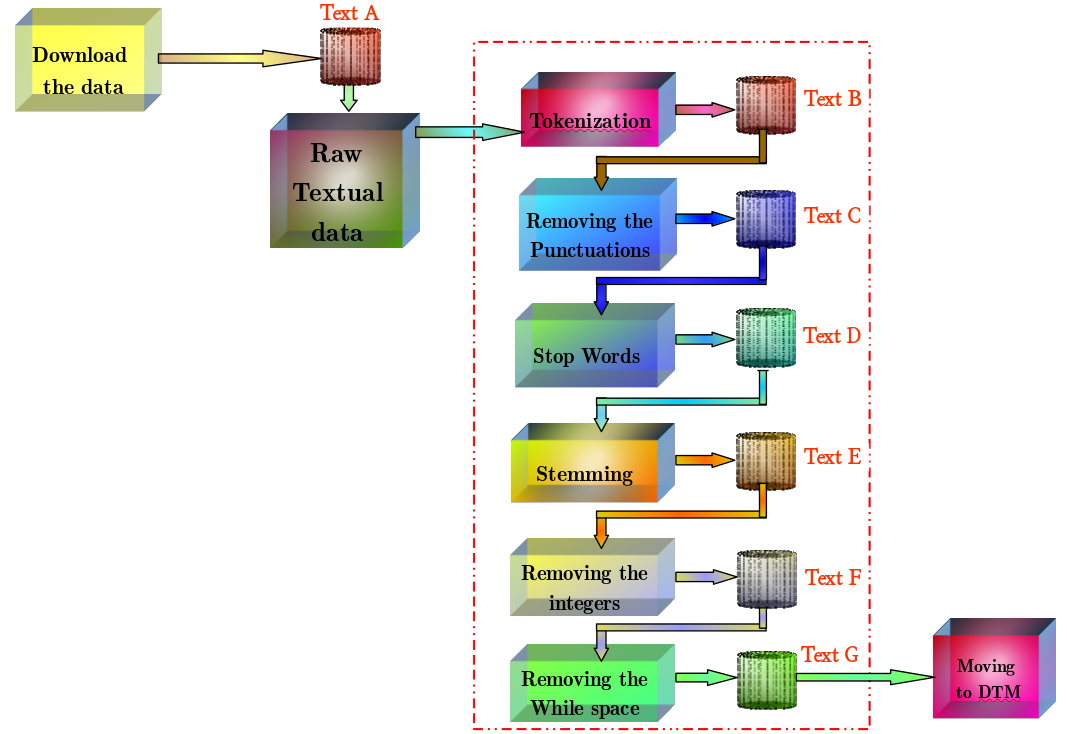}
	\caption{The different step of Pre-processing.}
	\label{fig:pre}
\end{figure}

It is also relevant to mention that by using text pre-processing, we can manage to extract the knowledge from the text data (corpus) which is needed for a good analysis and for an high accuracy of classifiers. We can see those steps in figure \ref{fig:pre}

\newpage
\subsection{Document term matrix}

	To make the unstructured texts data useful, we need to transform the textual data into vector spaces, which can be done by using what we call Bag-Of-Words (BOW) representation. The BOW converts the text data in a matrix format which simplify the statistical analysis of the data. The obtained matrix is called Document Term Matrix (DTM). The rows in this matrix represent the documents and the columns represent the terms. The DTM gives the occurrence of each term varying among the documents.

	In this study, we will use nine different books (T = 9), with 704 chapters/documents (n = 704) and 5131 (p=5131)  tokens/words. Hence, we have a 704 x 5131 DTM. When we consider the $t^{th}$ sacred scripture separately by means of its own DTM $X^{t}$, its representation looks like follows:

	\begin{equation} X^{(t)} = 
		\begin{bmatrix}
		X^{(t)}_{1,1} & X^{(t)}_{1,2} & \dots & \dots & \dots & X^{(t)}_{1,j_{t}}&\dots & X^{(t)}_{1,p_{t}}\\ \\
		X^{(t)}_{i_{t},1} & X^{(t)}_{i_{t},2} & \dots & \dots & \dots & X^{(t)}_{i_{t},j_{t}} &\dots& X^{(t)}_{i_{t},p_{t}} \\ \\
		\colon & \colon & \ddots &\ddots & \dots & \colon & \dots & \colon\\ \\
		X^{(t)}_{n_{t},1} & X^{(t)}_{n_{t},2} & \dots &\dots &\dots & 	X^{(t)}_{n_{t},j_{t}} &\dots & X^{(t)}_{n_{t},p_{t}}
		\end{bmatrix}
	\end{equation}

From this matrix, each column $X^{(t)}_{.,j_{t}}$ represents a term and each row $X^{(t)}_{i_{t},.}$ represents a document/chapter.

If we consider the whole corpus of sacred books, the Document Term Matrix is defined by the following relation:

	\begin{equation} X = 
		\begin{bmatrix}
		X^{(1)}_{1,1} & X^{(1)}_{1,2} & \dots & \dots & \dots & X^{(1)}_{1,j_{1}}&\dots & X^{(1)}_{1,p}\\ 
		\colon & \colon & \ddots &\ddots & \dots & \colon & \dots & \colon\\ 
		X^{(1)}_{i_{1},1} & X^{(1)}_{i_{1},2} & \dots & \dots & \dots & X^{(1)}_{i_{1},j1} &\dots& X^{(1)}_{i_{1},p} \\ 
		\colon & \colon & \ddots &\ddots & \dots & \colon & \dots & \colon\\ 
		X^{(1)}_{n_{1},1} & X^{(1)}_{n_{1},2} & \dots &\dots &\dots & X^{(1)}_{n_{1},J1} &\dots & X^{(1)}_{n_{1},p}\\ \\
		\colon & \colon & \ddots &\ddots & \dots & \colon & \dots & \colon\\ \\
		\colon & \colon & \ddots &\ddots & \dots & \colon & \dots & \colon\\ 
		X^{(t)}_{1,1} & X^{(t)}_{1,2} & \dots & \dots & \dots & X^{(t)}_{1,j_{t}}&\dots & X^{(t)}_{1,p}\\ 
		\colon & \colon & \ddots &\ddots & \dots & \colon & \dots & \colon\\ 
		X^{(t)}_{i_{t},1} & X^{(t)}_{i_{t},2} & \dots & \dots & \dots & X^{(t)}_{i_{t},j_{t}} &\dots& X^{(t)}_{i_{t},p} \\ 
		\colon & \colon & \ddots &\ddots & \dots & \colon & \dots & \colon\\ 
		X^{(t)}_{n_{t},1} & X^{(t)}_{n_{t},2} & \dots &\dots &\dots & X^{(t)}_{n_{t},j_{t}} &\dots & X^{(t)}_{n_{t},p}\\ \\
		\colon & \colon & \ddots &\ddots & \dots & \colon & \dots & \colon\\ \\
		\colon & \colon & \ddots &\ddots & \dots & \colon & \dots & \colon\\ 
		X^{(T)}_{1,1} & X^{(T)}_{1,2} & \dots & \dots & \dots & X^{(T)}_{1,j_{T}}&\dots & X^{(T)}_{1,p}\\ 
		\colon & \colon & \ddots &\ddots & \dots & \colon & \dots & \colon\\ 
		X^{(T)}_{i_{T},1} & X^{(T)}_{i_{T},2} & \dots & \dots & \dots & X^{(T)}_{i_{T},j_{T}} &\dots& X^{(T)}_{i_{T},p} \\ 
		\colon & \colon & \ddots &\ddots & \dots & \colon & \dots & \colon\\ 
		X^{(T)}_{n_{T},1} & X^{(T)}_{n_{T},2} & \dots &\dots &\dots & X^{(T)}_{n_{T},j_{T}} &\dots & X^{(T)}_{n_{T},p}\\ \\
		\end{bmatrix}
	\end{equation}

\subsection{Similarities between sacred texts}
	After converting the documents into term vectors, the similarity between them can be estimated by comparing the corresponding vectors. To distinguish the documents precisely, there are various algorithms which can be used. In this part, we will describe some of the standard distances used in text analysis like the cosine similarity, Euclidean distance, Manhattan distance \citep{qahl2014automatic}.

\subsubsection{Term frequency inverse document frequency}

	The most common document similarity algorithm is term frequency-inverse document frequency (Tf-IDF). he Term Frequency (TF) means that the more frequently one term appears in a document the more its weight increases. In inverse document frequency (IDF), the term occuring in a greater number of documents are relatively less relevant and should be weighted less. It is presented mathematically by the equation~\ref{TF} where $freq_{i,j}$ represents the number of occurrences of the word j in file i and $W_{ij}$ is the weight of word j in file i.
	
	\begin{eqnarray}
		tf_{i,j} = \dfrac{freq_{i,j}}{\max(freq_{i,j})}
		\label{TF}
	\end{eqnarray}

From equation~\ref{TF1},  $m$ represents the number of files, $m_{j}$ represents the number of files containing the word j.

	\begin{eqnarray}
		idf_{i,j} = \log\dfrac{m}{m_{j}}
		\label{TF1}
	\end{eqnarray}

And the weight is calculated according to equation~\ref{TF2}:
	\begin{eqnarray}
		W_{i,j} = tf_{i,j}\cdot idf_{i,j} 
		\label{TF2}
	\end{eqnarray}

\subsubsection{Distance measures}

	In this part of our study, we will explore the similarity respectively dissimilarity between chapters and documents by applying the non-standard distances like cosine similarity and also some standard distances such as Euclidean distance, Manhattan distance and Jaccard similarity \citep{Fok2019asian}. Notice that not every distance measure is a metric. A metric must satisfy four conditions. Let $A$ and $B$ be any two documents and $d(A , B)$ be the distance between both \citep{maher2016effectiveness}. The conditions read as follows:

\begin{itemize}
	\item The distance between any two given points must be
	nonnegative. Mathematically, it is:
	$$ d(A , B) \geq 0 $$
	
	\item The distance between two points is equal to zero (0) if and only if the two points are the same. It is formulated:
	$$ d(A , B) = 0  \hspace{0.5cm} iff \hspace{0.5cm} A = B $$
	
	\item The distance must be symmetric, which means that the distance from A to B is the same as the distance from B to A.
	$$ d(A ,B) = d(B ,A) $$
	
	\item The distance measure must satisfy the inequality~\ref{equ} which is the triangle inequality.
	\begin{eqnarray}
	d(A , C) \leq d(A , B) + d(B , C)
	\label{equ}
	\end{eqnarray} 
\end{itemize}

The different distance similarity measures which can  facilitate the interpretation or understanding of the similarity respectively dissimilarity between documents can also be applied for the corpus. In text mining, we deal with high-dimensional data. The sparsity of the raw data makes the study more complex.

\begin{itemize}
	\item The Euclidean distance measure :
	
	The euclidean distance can be defined as the shortest straight-line distance between two points. It is part of the Minkowski family. The Minkowski distance is a metric distance class on the Euclidean space \citep{deza2006dictionary}. The mathematical representation of Eucleidean distance is given by equation~\ref{Eu}.
	\begin{eqnarray}
		d_{E}(X_{l}^{(a)} , X_{m}^{(b)}) = \sqrt{\sum_{j = 1}^{p} ( X_{l,j}^{(a)} - X_{m,j}^{(b)} )^{2}}
		\label{Eu} 
	\end{eqnarray}
	
	\item The Jaccard similarity measure:
	
	The Jaccard similarity measure is approached used to measure the similarity between two chapters/documents by taking the intersection of both and divide it by their union \citep{zahrotun2016comparison}. The Jaccard coefficient between two documents $X_{l}^{(a)}$ and $X_{m}^{(b)}$ is mathematically defined by: 
	\begin{eqnarray}
	sim(X_{l}^{(a)} , X_{m}^{(b)}) \equiv sim(X_{l} , X_{m}) = \dfrac{\displaystyle\sum_{j=1}^{p}\min \{X_{lj},X_{mj}\}}{\displaystyle\sum_{k=1}^{p}\max \{X_{lk},X_{mk}\}}
	\label{Jac} 
	\end{eqnarray}
	
	Thus, by using the equation \ref{Jac} the Jaccard distance between two document is given by the equation~\ref{Jac2}:
	\begin{eqnarray}
	d_{J}(X_{l}^{(a)} , X_{m}^{(b)}) \equiv d_{J}(X_{l} , X_{m}) =1 - sim(X_{l} , X_{m})
	\label{Jac2} 
	\end{eqnarray}

	\item The Manhattan distance measure :
	
	The Manhattan Distance is the sum of absolute differences between points across all the dimensions.
	\begin{eqnarray*}
		d_{M}(X_{l}^{(a)} , X_{m}^{(b)}) = \sum_{j = 1}^{p} | X_{l,j}^{(a)} - X_{m,j}^{(b)}|
	\end{eqnarray*}

	\item The Cosine similarity measure 
	
	The Cosine similarity measure considers the correlation between the vectors. It is also the most popular similarity measure applied to text documents \citep{maher2016effectiveness}. The formulation of this measure is defined by:
	\begin{eqnarray}
	d_{Cos}(X_{l}^{(a)} , X_{m}^{(b)}) \equiv d_{Cos}(X_{l} , X_{m}) =  \dfrac{X_{l}^{T}X_{m}}{(X_{l}^{T}X_{l})^{\frac{1}{2}}(X_{m}^{T}X_{m})^{\frac{1}{2}}}
	\label{Jac2} 
	\end{eqnarray}
	
\end{itemize}

\subsubsection{Text similarity measures}

After having introduced the distance similarity measures, we can use them to explore the similarity between books respectively documents.

First of all, we are going to compare the similarity between chapters within the same book. For instance in book $X_{\alpha}$, the distance between the chapters is mathematically represented by:
	\begin{eqnarray*} D^{(\alpha)} = 
		\begin{bmatrix}
			d^{(\alpha)}_{1,1} & d^{(\alpha)}_{1,2} & \dots & d^{(\alpha)}_{1,n_{\alpha}}\\ \\
			d^{(\alpha)}_{2,1} & d^{(\alpha)}_{2,2} & \dots & d^{(\alpha)}_{2,n_{\alpha}}\\ \\
			\colon & \colon & \ddots & \colon\\ \\
			d^{(\alpha)}_{n_{\alpha},1} & d^{(\alpha)}_{n_{\alpha},2} & \dots & d^{(\alpha)}_{n_{\alpha},n_{\alpha}}\\ \\
		\end{bmatrix}
	\end{eqnarray*} 

Notice that the matrix $D^{(\alpha)}$ is  nonnegative $n_{\alpha} \times n_{\alpha}$  $\left ( \mathbb{R}^{n_{\alpha} \times n_{\alpha}}_{+} \right )$ .

The components are defined as :
$ d^{(\alpha)}_{l , m} \equiv d(X^{(\alpha)}_{l} , X^{(\alpha)}_{m}) \equiv $ the distance between the $l^{th}$ chapter and the $m^{th}$ chapter of the $\alpha^{th}$ book $X^{(\alpha)}$. This can help us to evaluate the relationship between various chapter within the same book.

If we want to discover the relationship between books $X_{\alpha}$ and $X_{\beta}$, we can use the following formula:
	\begin{eqnarray*} D= 
		\begin{bmatrix}
			d_{1,1} & d_{1,2} & \dots & d_{1,n}\\ \\
			d_{2,1} & d_{2,2} & \dots & d_{2,n}\\ \\
			\colon & \colon & \ddots & \colon\\ \\
			d_{n,1} & d_{n,2} & \dots & d_{n,n}\\ \\
		\end{bmatrix}
	\end{eqnarray*} 

Also this $n \times n$ matrix is  nonnegative $( \mathbb{R}^{n \times n}_{+})$.

$ d_{l , m} \equiv d(X^{(\alpha)}_{l} , X^{(\beta)}_{m}) \equiv $ is the distance between the $l^{th}$ chapter of the book $X^{(\alpha)}$ and the $m^{th}$ chapter of book $X^{(\beta)}$. This can help us to evaluate the relationship between various chapter within the same book.

There are four different approaches existing to study the similarity between the sacred texts reading as following:

\begin{enumerate}
	\item The first approach for obtaining the distance between books is named single or minimum linkage and is given by the relation \ref{min}. This simply means that the distance between two books, for instance $X^{(\alpha)}$ and $X^{(\beta)}$, corresponds to the smallest value of the $n * n$  distance between their chapters. Mathematically, it is define as:
	\begin{eqnarray}
		\Delta (X^{(\alpha)} , X^{(\beta)}) = \min_{\stackrel{l\in[n_{\alpha}]}{m\in[n_{\beta}]}}\{d(X^{(\alpha)}_{l} , X_{m}^{(\beta)})\}
		\label{min}
	\end{eqnarray}
	
	\item The second approach for measuring distances between books is the maximum linkage given by the relation \ref{max}. In this method the distance between two books, for instance $X^{(\alpha)}$ and $X^{(\beta)}$, corresponds to the largest value of the $n * n$  distance matrix with respect to their chapters. Mathematically, it is defined as:
	\begin{eqnarray}
		\Delta (X^{(\alpha)} , X^{(\beta)}) = \max_{\stackrel{l\in[n_{\alpha}]}{m\in[n_{\beta}]}}\{d(X^{(\alpha)}_{l} , X_{m}^{(\beta)})\}
		\label{max}
	\end{eqnarray}
	
	\item The third approach for the distance between books is the average linkage and is defined by the relation \ref{mean}. Here, the distance between two books, for instance $X^{(\alpha)}$ and $X^{(\beta)}$, corresponds to the mean value of the $n * n$  distance matrix with respect to their chapters. It is written as:
	\begin{eqnarray}
		\Delta (X^{(\alpha)} , X^{(\beta)}) = mean_{\stackrel{l\in[n_{\alpha}]}{m\in[n_{\beta}]}}\{d(X^{(\alpha)}_{l} , X_{m}^{(\beta)})\}
		\label{mean}
	\end{eqnarray}
	
	\item The last approach for distance between books is the median distance and it is defined by the relation \ref{men} which reads:
	\begin{eqnarray}
		\Delta (X^{(\alpha)} , X^{(\beta)}) = median_{\stackrel{l\in[n_{\alpha}]}{m\in[n_{\beta}]}}\{d(X^{(\alpha)}_{l} , X_{m}^{(\beta)})\}
		\label{men}
	\end{eqnarray}
\end{enumerate} 

Having $B$ sacred books, we can give the expression of the distance between the whole set of book as
	\begin{eqnarray*} \Delta= 
		\begin{bmatrix}
			\Delta_{1,1} & \Delta_{1,2} & \dots & \Delta_{1,B}\\ \\
			\Delta_{2,1} & \Delta_{2,2} & \dots & \Delta_{2,B}\\ \\
			\colon & \colon & \ddots & \colon\\ \\
			\Delta_{B,1} & \Delta_{B,2} & \dots & \Delta_{B,B}\\ \\
		\end{bmatrix}
	\end{eqnarray*} 

where $ \Delta_{\alpha,\beta} \equiv \Delta (X^{(\alpha)}, X^{(\beta)})$ corresponds to the distance to the book $\alpha$ and the book $\beta$.

\subsection{Learning sacred text recognition} 

	For this research work, we use the Asian scripture, the Bible and Quran as our collection of texts (data), also named corpus in this context. When we want to obtain the origin of a text or of a text fraction, we can use machine learning making predictions about the source of these texts. Thus, in what follows, we will describe unsupervised and supervised machine learning applied to text mining processes.

\subsubsection{Introduction to machine learning}

	In machine learning, we have data with which the machine is trained. Based on this training, it learns patterns in the data. For text analytics, machine learning involves a set of statistical techniques for identifying various characteristics and patterns among the considered texts. Machine learning is divided into supervised and unsupervised learning which will be described in the following. Due to the fact that the text data can have hundreds of thousands of dimensions (sentences or words), text data requires a particular approach to machine learning.

\begin{itemize}
	\item \textbf{Unsupervised learning}:

	Unsupervised machine learning algorithms infer patterns from a dataset without reference to known or labeled outcomes. Unlike supervised machine learning, unsupervised machine learning methods cannot be directly applied to a regression or a classification problem because of the missing labels. It is about training a model without pre-tagging or annotating.

	One of the most interesting questions is whether it is possible to classify texts according to the sacred books that they originate from.  In this section, as far as the relationships among entire sacred scriptures are concerned, we herein briefly describe the way in which we use cluster analysis to tackle and answer that overarching question.

	\item \textbf{Supervised learning}:
	
	In supervised learning the goal is to predict the label of a data sample. In the context of sacred text classification, we use labelled documents to train a model. In supervised machine algorithm, the goal is to achieve high prediction accuracy after training with the machine. 
	
	So far, there exists a couple of algorithms that we can apply on the labeled corpus. Inter alia, we have the linear classification algorithms like K-nearest neighbors (\textbf{KNN}), Random Forest (\textbf{RF}) and we also have the Support Vector Machines (\textbf{SVM}) which can be both linear and non-linear.

	\begin{itemize}
		
		\item The K-Nearest Neighbors (KNN):
		
		In order to proceed with the categorization of texts, there are a number of methods and techniques that we can apply. Seeing a new document, the KNN method assigns to this document the label that occurs the most often among its closest neighbors.

		\item The Support Vector Machine (SVM):
		
		The support vector machine (SVM) is one of the methods of supervised learning that generates input-output mapping functions from a set of labeled training data \citep{wang2005support}.
		To perform classification, the input data is projected to an higher dimensional space feature space through nonlinear kernel functions enforcing the seperability of the input data. In addition, the SVM creates a margin between the separating hyperplane and the data.

		The equation of the hyperplane is mathematically defined as follow:
		$$ w^{T}x - b = 0 $$
		
		The equation of separating hyperplanes are given by:
		\begin{eqnarray}
			\begin{cases}
				w^{T}x - b = 1\\
				w^{T}x - b = -1
			\end{cases}
		\end{eqnarray} 
		
		The first instance is about the \textbf{Hard Margin} and the goal here is to maximize $ \dfrac{2}{\parallel W \parallel} $ which is equivalent to minimizing $ \dfrac{\parallel w \parallel}{2} $.
		
		We must have two (2) constraints:
		\begin{eqnarray}
			\begin{cases}
				w^{T}x^{(i)} - b \geq 1  \hspace{0.8cm} if \hspace{0.8cm} y^{(i)} = 1\\
				w^{T}x^{(i)}  - b \leq -1 \hspace{0.8cm} if \hspace{0.8cm} y^{(i)} = -1
			\end{cases}
		\end{eqnarray}
		
		By combination of constraint, we have the following:
		\begin{center}
			$ y^{(i)}(w^{T}x^{(i)} - b) \geq 1 $ for $i = 1 ,..., m$
		\end{center}
		
		For this second instance, the optimization function is defined as:
		$$ w^{*} = argmin \parallel w \parallel^{2} $$
		
		The second instance is the \textbf{Soft Margin}. It allows for points to be inside the margin. The loss function becomes the hinge loss which is used when one deals with discrete labels. It is mathematically defined by:
		\begin{eqnarray}
			\mathcal{L}(w) = \dfrac{1}{m}\sum_{i=1}^{m} l(w; x,y) + \lambda \parallel W \parallel^{2}_{2}
		\end{eqnarray}
		
		with  $$ l(w; x,y)  = max[0 , 1 - y^{(i)}(w^{T}x^{(i)} - b) ]$$
		
		Here, the optimization function is define as: $ w^{*} = argmin \mathcal{L}(w)   $
		
		The third instance is the \textbf{non-linear classification}. In this case, the feature $x^{i}$ moves to the feature space $\phi^{i}$ and the kernel trick is given by:
		$$ K(x, x') = \phi^{T}(x) \cdot \phi(x') $$
		
		And the weight is equal to:
		\begin{eqnarray*}
			w &=& \sum_{i=1}^{m} \alpha_{i} y^{(i)} \phi(x^{(i)})\\
			w^{T} &=& \sum_{i=1}^{m} \alpha_{i} y^{(i)} \phi^{T}(x^{(i)})\\
			w^{T} \phi(x^{(i)}) &=&  \sum_{i=1}^{m} \alpha_{i} y^{(i)}K(x^{(i)}, x')
		\end{eqnarray*}

		\item The Random Forest (RF):
		
		We can name Random Forest as another tree classifier, which can be used for classification of text data. Because of the fact that it is capable of extracting similarity between features, it is also used in text mining as an embedded feature selection method. 
		
	\end{itemize}
\end{itemize}

\subsubsection{Classification function}

	Let $x$ be the vector of word frequency corresponding to a chapter from one of the nine sacred texts. The probability that this chapter $x$ comes from the $t^{th}$ book is denoted as
	$\mathbb{P} [y = s_{t}| x] $ where $y$ is the variable indicating the sacred book.

The function $f$ according to which classification is done reads:
	\begin{eqnarray}
		f(x) = argmax_{t\in[T]}\{\mathbb{P}[y = s_{t}| x]\} 
	\end{eqnarray}
	For the pure purpose of scripture authentication, we build the classifier from the data $\hat{f}(x)$. For instance, the classifier function for K-Nearest-Neighbors is given by:
	$$ \hat{f}^{(KNN)}(x) = argmax_{t\in T} \left\{\dfrac{1}{k} \sum_{j=1}^{n} \mathbb{I}(y_{j} = t) \mathbb{I}(x_{j} \in \mathcal{V}_{k}(x))\right\}  $$

\subsubsection{Cross validation and Grid searching}

	To evaluate the performance of our model, we will  Cross validate. Another method that is used for the selection of the best hyperparameter configuration is the Grid search.

\begin{itemize}
	\item m-Fold Cross validation
	
	Inspecting the test error, one can separate the data into a training and test set. However, this process may lead accuracies that are very different from one test to another using the same algorithm. To avoid this issue, we will use the \textbf{m-Fold Cross-Validation} where $m$ is a indicating the number of folds. In this process, we divide the data into $m$ folds. A training set is represented by $m-1$ set while the testing set is the remaining. After each iteration it shuffle the data and create a new training and testing sets.
	
	The cross-validation estimate of prediction error is given by
	$$ CV(\hat{f}) = \dfrac{1}{N}\sum_{i=1}^{N}L(y_{i} , \hat{f}^{-k(i)}(x_{i})) $$
	
	where  $k(i)$ means, the model f is trained without the training patterns in the same partition of the dataset as pattern $i$.
	
	\item Grid searching
	
	The hyperparameter optimization is essential in machine learning due to the fact that neural networks are difficult to configure and there are a lot of parameters that need to be set. Setting the hyperparameters manually can consume a lot of time. Hence, we will use the grid search method for this purpose.
	
	Grid-searching is the process of scanning the data to configure optimal parameters for a given model. Depending on the type of model that is used, we need to adjust the parameters of the grid search algorithm. Grid-searching stores a model by doing the iteration through every parameter combination and selects the optimal configuration for our model. The general formulation to compute the accuracy is given by:
	\begin{eqnarray}
	Accuracy = \dfrac{TP + TN}{TP + TN + FP + FN}
	\end{eqnarray}
\end{itemize}

Where TP, TN, FP, FN means True Positive,  True Negative, False Positive, False Negative respectively.

\section{Results}

	We are presenting the different results we got during our analysis. Starting by describing our data, continuous by the distribution of the words, then we perform some text analysis and end by presenting the distance measurement.

\subsection{Description of the data}

	Before starting the pre-processing, we needed to download the English version of Quran. Then we will do the analysis on our data and transform the text into tokens. This transformation yields  in total 156110 tokens. This step is followed by removing useless tokens. Hence, punctuations ( e.g.   - , : ""), stop words (e.g. as, it, very, own, any, only, off) and integers (e.g. 1, 2, 3 etc) are removed. Finally, we get the cleaner data ready to be used.

\subsection{Distribution of words}

	The frequency distribution contains the ratios between each token and the total number of tokens shown below in figure \ref{fig:tex1}.

	\begin{figure}[htbp!] 
		\centering
		\includegraphics[height =9cm, width=12cm]{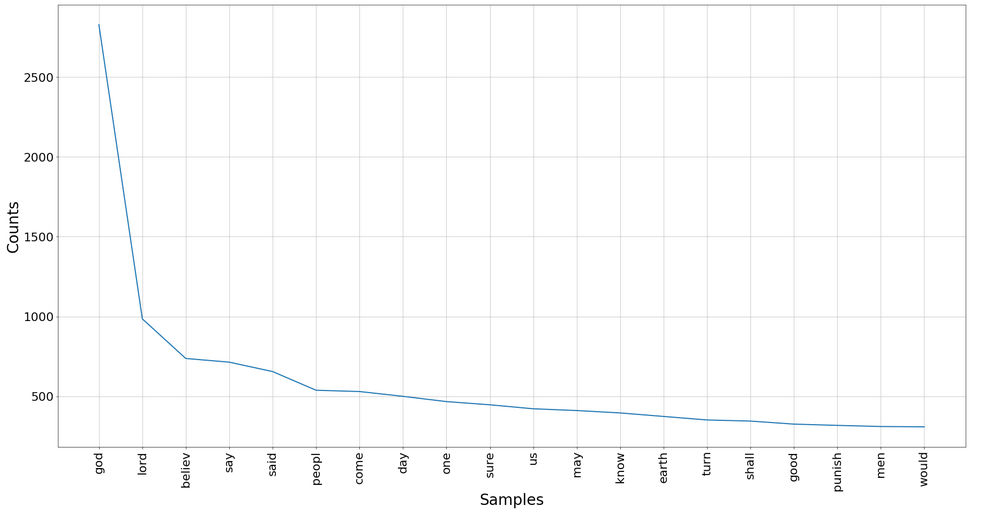}
		\caption{Frequency distribution of words in text data.}
		\label{fig:tex1}
	\end{figure}

	In this graph, we can see that the word \textbf{god} is the most often used word. It is followed by the words \textbf{lord} and \textbf{belive}.

\subsection{Frequency of Part-Of-Speech}

	In this section, we will present the Part-Of-Speech tagging that, we get from the list of tokens. Then, we will also present the frequency plot of those POS tagging.
\subsubsection{The Part Of Speech tagging}
	The following table, shows the few example of POS tagging which are from our list of tokens.

\begin{table}[htbp]
	\centering
	\begin{tabular}{|l|l|}
		\hline 
		\textbf{Object} & \textbf{Label prefix} \\ 
		\hline 
		god & \texttt{NN} \\ 
		\hline 
		lord & \texttt{NN} \\ 
		\hline 
		said & \texttt{VBD} \\ 
		\hline 
		believ & \texttt{NN} \\ 
		\hline 
		allah & \texttt{NN} \\ 
		\hline 
		may & \texttt{MD} \\
		\hline 
	\end{tabular} \\
	\caption{The Part Of Speech tagging for the data}
\end{table}

	\begin{figure}[htbp!] 
		\centering
		\includegraphics[height =5.5cm, width=8.5cm]{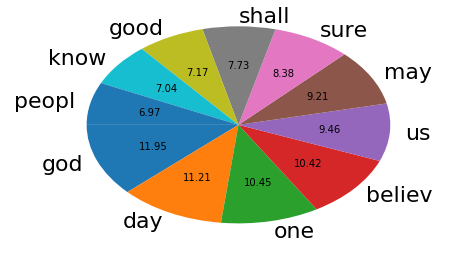}
		\caption{Frequency of POS tagging in text data.}
		\label{fig:tex}
	\end{figure}

From subsection 3.1 to session 3.3, we repeat the same method for the Asian books and the Bible.
\subsection{Performance analysis}

\subsubsection{Document Term Matrix}

\begin{itemize}
	\item Document Terms Matrix of Quran
	
	The following table describes the labeled document term matrix of Quran. From this table, we obtain that the sacred book has 114 chapters/documents and 4284 terms/tokens.  The value in each cell represents the number of times that the word corresponding to that column occurs in the documents corresponding to that row.
	
	\begin{figure} 
		\centering
		\includegraphics[height=8cm, width=15cm]{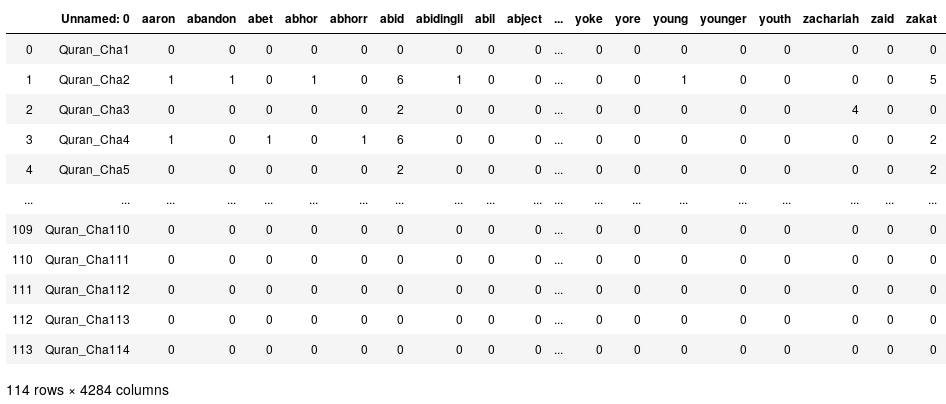}
		\label{fig:SM}
	\end{figure}

	\newpage	
	\item Document Terms Matrix of Asian and Bible
	
	The following table describe the labeled  document term matrix of Asian and The holy Bible. The table shows that these sacred books have together 590 chapters/documents and 1024 terms/tokens.

	\begin{figure}[htbp!] 
		\centering
		\includegraphics[height=7cm, width=14cm]{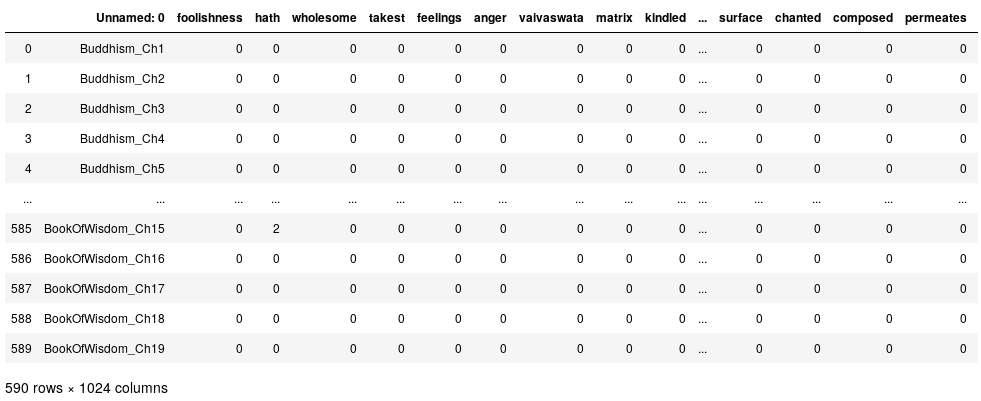}
		\label{fig:SM}
	\end{figure}
	
	\item Document Terms Matrix of both documents.
	
	Firstly, when we put all the sacred texts together, we get the DTM with 704 chapters/documents and 5131 terms/tokens. We notice that there are some missing values which will not be needed for our analysis. Thus, the next step consists of removing them. After removing all the missing values, we get the table below. It still has 704 chapters/documents and 5131 terms/tokens. This DTM will be used for several analysis that we make on the sacred scriptures.
	
	\begin{figure}[htbp!] 
		\centering
		\includegraphics[height=6.5cm, width=14cm]{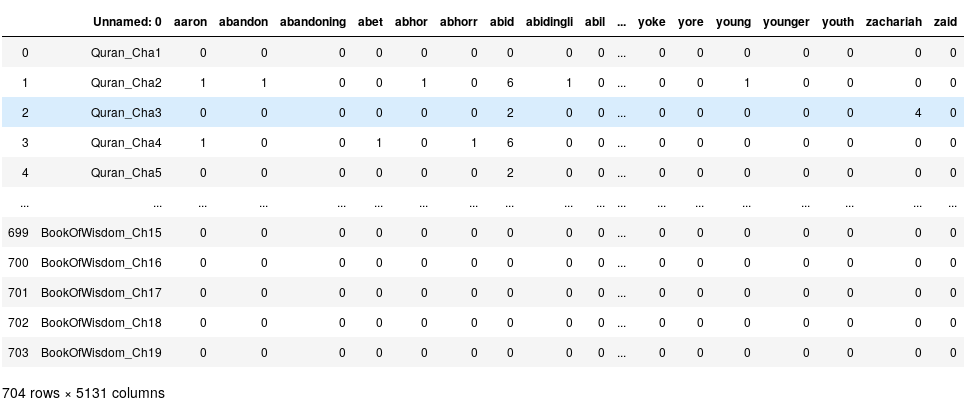}
		\label{fig:SM}
	\end{figure}
	
	\newpage
	\par Since we already have the matrix representation of the different sacred texts, corresponding here to the previous document term matrix (DTM), we will proceed with the studies of the distances/similarities between the chapters of the same book, between the chapters of a sacred scripture with those of other sacred books and finally between the different books without dividing them into chapters.
	
\end{itemize}

\subsection{Distance Measure}

	After converting the sacred books into term vectors, the distances/similarities between two documents can be estimated by comparing the corresponding vectors. In this part, we will apply the Euclidean distance, the Manhattan distance, the cosine similarity and the Jaccard distance.

\subsubsection{Distance between chapters of all sacred books}
	\begin{itemize}
		\item \textbf{Euclidean distance}
	\end{itemize}

Using the Euclidean distance, we can visualize the distance matrix by its corresponding heatmap.

\begin{figure}[htbp!]
	\centering
	\label{fig:examplePic} 
	\includegraphics[width=11cm, height = 8cm]{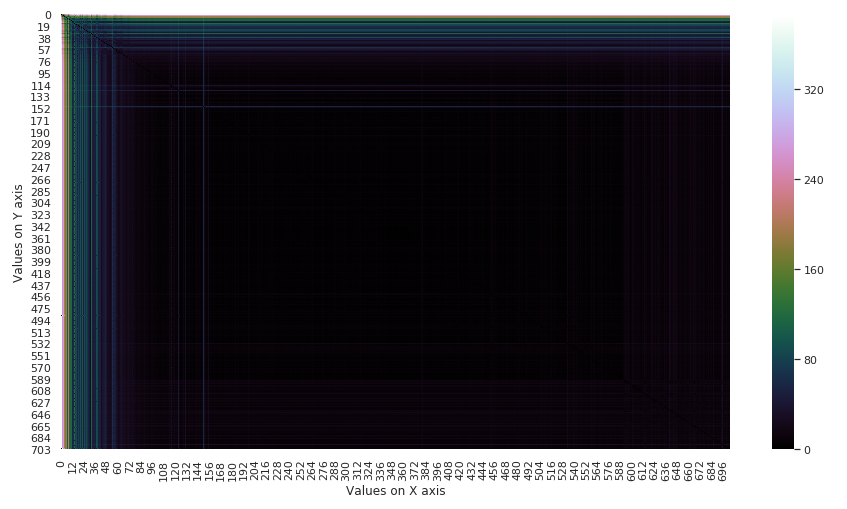}
	\caption{Heatmap of the Euclidean distance between the chapters in both sacred books.}
\end{figure}
	This heat map shows for instance that there is a very high distance between $10-th$ document corresponding to chapter 9 of Quran and the document with index $600$ corresponding to chapter 14 in the Book of Wisdom.

\begin{itemize}
	\item \textbf{Cosine distance}

	Analogously, we obtain a distance matrix after applying the cosine distance and its heatmap representation depicted below. 
	
	\begin{figure}[htbp!] 
		\centering
		\includegraphics[height=7.5cm, width=12cm]{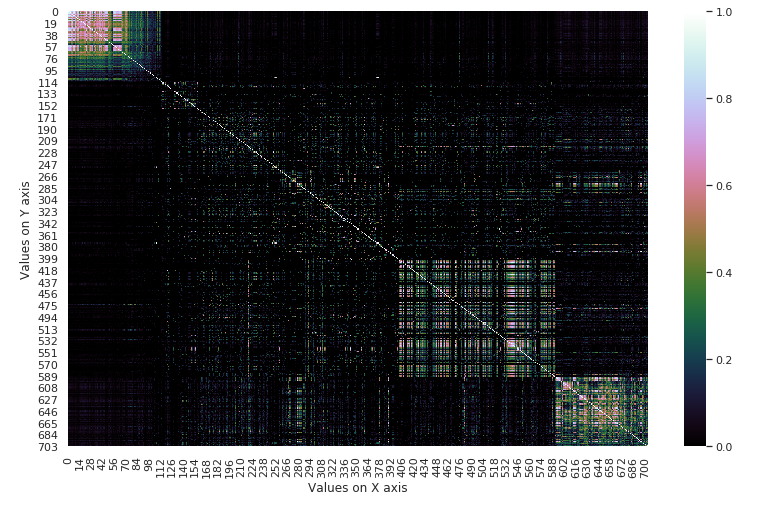}
		\caption{Heatmap of the cosine distance between the chapters in both sacred books.}
		\label{fig:SM}
	\end{figure}
	
	This heat map shows for instance that, there is a considerable distance between the chapters of the books. This is the reason why we have several block of color in the heat map.
\end{itemize}

\subsubsection{Distances measure between sacred books}

Regarding this part, we will show the different distance measure between the sacred books. 
\begin{itemize}
	\item \textbf{Euclidean distances}
	
	\begin{figure}[htbp!] 
		\centering
		\includegraphics[height=6.5cm, width=11cm]{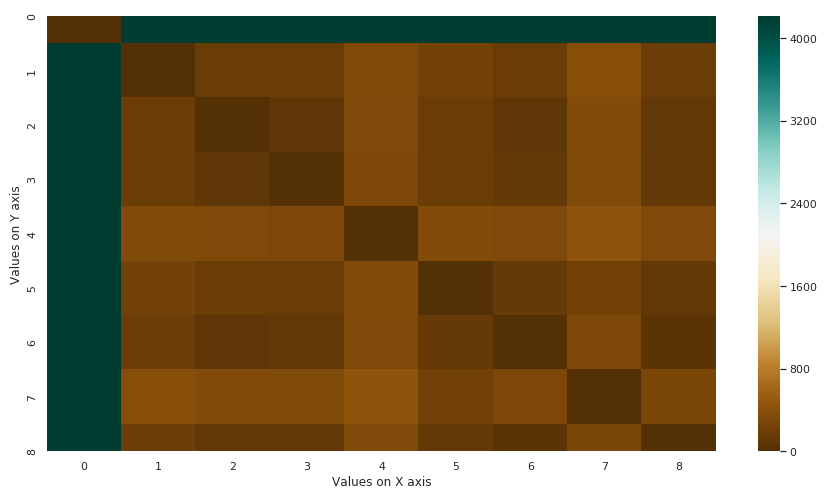}
		\caption{Heatmap of the Euclidean distance between sacred books}
		\label{fig:SM3}
	\end{figure}
	\begin{tabular}{|c|c|c|c|c|c|c|}\hline
		\textbf{Labels} & 0 & 1 & 2 & 3 & 4 & 5 \\ \hline
		\textbf{Books} & \textbf{Quran} & \textbf{Buddhism} & \textbf{Tao-Te-Ching} & \textbf{Upanishad}  & \textbf{Yogasutra} & \textbf{Book of Proverb}  \\\hline
	\end{tabular}
	
	\begin{tabular}{|c|c|c|}\hline
		6 & 7 & 8   \\ \hline
		\textbf{Book of Ecclesiasticus} & \textbf{Book of Ecclesiastes} & \textbf{Book of Wisdom}   \\\hline
	\end{tabular}

	From this heat map, we can say that there is a distance between the Quran and the other sacred books whereas there is a similarity between the book $6$ corresponding to book of book of ecclesiastes with the book $3$ corresponding to Tao Te Ching.
	
	\item \textbf{Cosine distance}
	
	\begin{figure}[htbp!] 
		\centering
		\includegraphics[height=7cm, width=11cm]{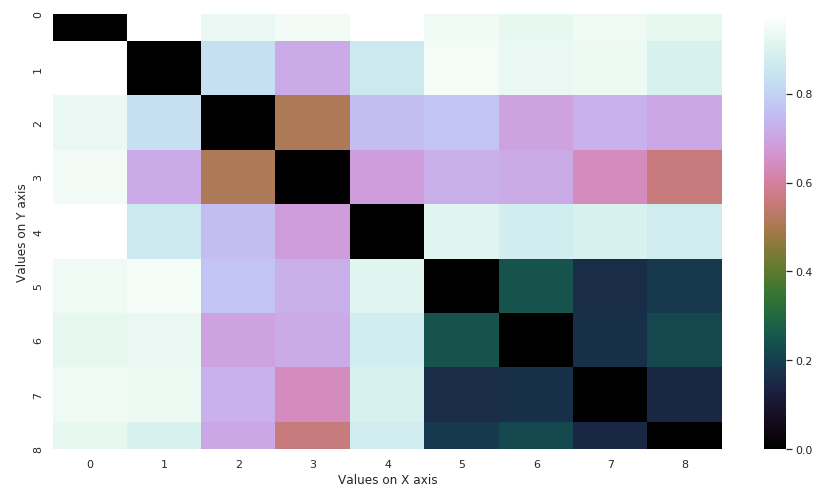}
		\caption{Heatmap of the cosine distance between the sacred books.}
		\label{fig:S8M}
	\end{figure}
\end{itemize}

	Here, the heat map shows that, still, there is a distance between the Quran and some of the other sacred books. We realize that, some of the Asian books (like Buddhism and YogaSutra) are not similar to the Bible.

\subsection{Correlation between the distance measures}

	In this section of our work, we will check if the different distance measures are based on the same properties. To do this, we will compute the correlation between the distance measures applied before.
\begin{figure}[htbp!] 
	\centering
	\includegraphics[height=4.5cm, width=8.5cm]{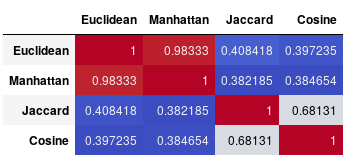}
	\caption{Correlation matrix between the values obtained from the different distance measures}
	\label{fig:oiio}
\end{figure}

\subsection{Classification of the fragment of sacred text}

	Here, we used four different model to do the prediction. We have, K-Nearest Neighbors (KNN), Super vector Machine (SVM), Random Forest (RF) and multinomial naive bayes. Grid-search is used to find the optimal hyper-parameters of a model which give the higher accuracy. We used m = 10 folds in this work.

	Thus, we find the accuracy for the RF equal to 0.7370, for the Super Vector Machine SVM  classifier is 0.7781, for the KNN find 0.6128 and 0.8584 for the Multinomial Naive Bayes  classifier
\begin{figure}[htbp!] 
	\centering
	\includegraphics[height=7cm, width=8.1cm]{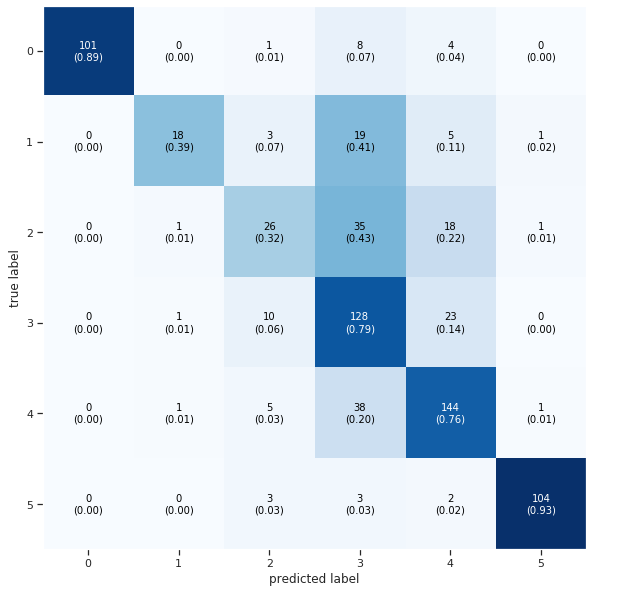}
	\includegraphics[height=7cm, width=8.1cm]{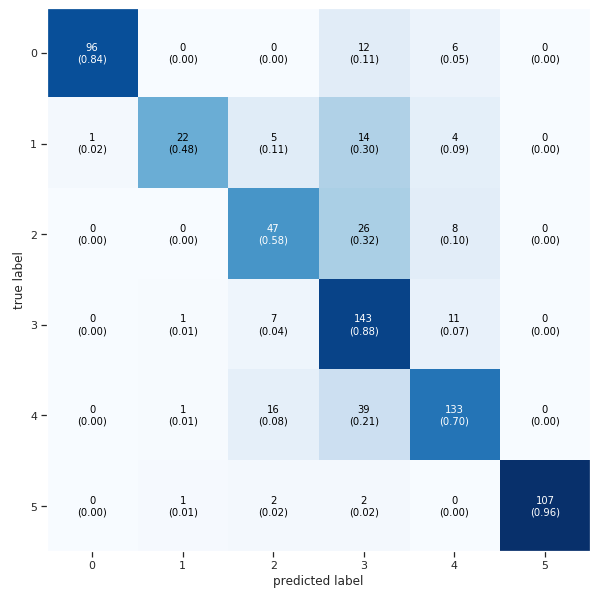}
	\caption{The confusion matrix of the Random Forest and Super Vector Machine}
	\label{fig:oi6io}
\end{figure}

\begin{figure}[htbp!] 
	\centering
	\includegraphics[height=7cm, width=8.1cm]{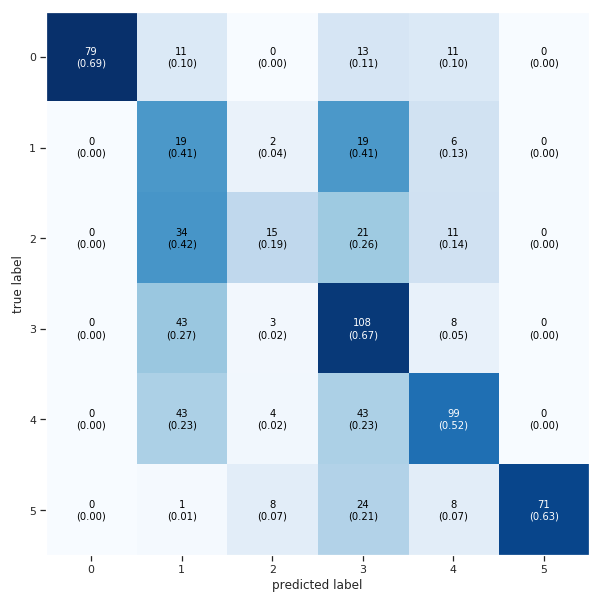}
	\includegraphics[height=7cm, width=8.1cm]{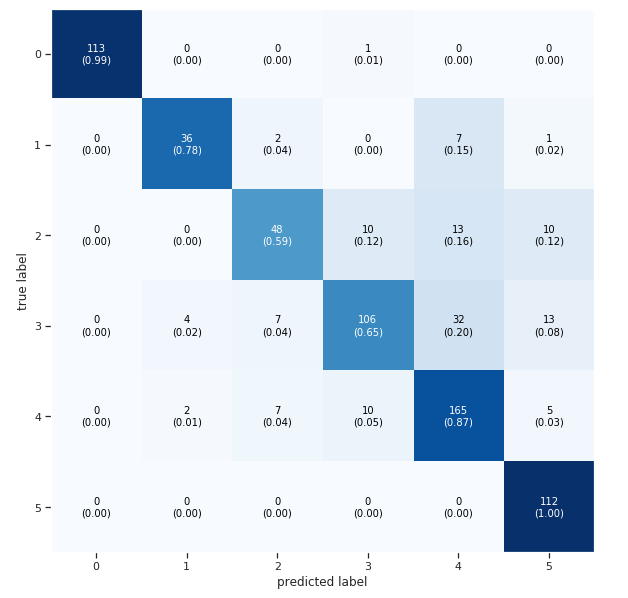}
	\caption{The confusion matrix of the K-Nearest Neighbors and Multinomial Naive Bayes}
	\label{fig:oii1o}
\end{figure}

\newpage

\section{Discussion}
When we carried out the natural language processing using the NLTK Library, we focused on three things that are essential in text mining: We applied pre-processing steps as well as the Part-Of-Speech tagging to obtain appropriate text data for the analysis. Finally, we worked out the document term matrix representation of the text data with respect to several distance measures.

Looking at figure \ref{fig:SM3}, the Euclidean measure tells us that the distance between the Quran and other sacred books seems to be high. The Asian books are almost similar to the sacred books from Bible. But looking at figure \ref{fig:S8M}, we notice that, the distance between the Quran and some sacred books is not very high. The books from the Bible and Asian religions are almost similar. This leads us to wonder why the distance measure methods do not give us the same result. This is the reason why it is  interesting to check the correlation between these approaches.

When we look at figure \ref{fig:oiio}, the Euclidean distance and Manhattan distance are strongly correlated while we do not observe a high correlation with the Jaccard and Cosine distance. Based on this result, we conclude that, not all of these methods use the same properties to determine the distance between documents.

To analyze whether the high similarities between documents obtained with respect to the different measures are related to similar meanings among these documents, we consider exemplifying the following fragments:

\textbf{1- Tao Te Ching :} \textit{ Not to value and employ men of superior ability is the way to keep the people from rivalry among themselves; not to prize articles which are difficult to procure is the way to keep them from becoming thieves; not to show them what is likely to excite their desires is the way to keep their minds from disorder. Therefore the sage, in the exercise of his government, empties their minds, fills their bellies, weakens their wills, and strengthens their bones. He constantly (tries to) keep them without knowledge and without desire, and where there are those who have knowledge, to keep them from presuming to act (on it). When there is this abstinence from action, good order is universal. } 

\textbf{2- Book of Wisdom :} \textit{ - the works of the hands of men.  13:1. But all men are vain, in whom there is not the knowledge of God: and who by these good things that are seen, could not understand him that is, neither by attending to the works have acknowledged who was the workman:  13:2. But have imagined either the fire, or the wind, or the swift air, or the circle of the stars, or the great water, or the sun and moon, to be the gods that rule the world.  13:3.}

\textbf{3- Tao Te Ching :} \textit{ - The skilful traveller leaves no traces of his wheels or   footsteps; the skilful speaker says nothing that can be found fault with or blamed; the skilful reckoner uses no tallies; the skilful   closer needs no bolts or bars, while to open what he has shut will be impossible; the skilful binder uses no strings or knots, while to unloose what he has bound will be impossible. In the same way the sage is always skilful at saving men, and so he does not cast away any man; he is always skilful at saving things, and so he does not cast away anything. This is called 'Hiding the light of his procedure.'Therefore the man of skill is a master (to be looked up to) by him who has not the skill; and he who has not the skill is the helper of (the reputation of) him who has the skill. If the one did not honour his master, and the other did not rejoice in his helper, an (observer), though intelligent, might greatly err about them. This is called 'The utmost degree of mystery.'}

\textbf{4-UgaSutra:} \textit{ -  The disciple said: I do not think I know It well, nor do I think that I do not know It.  He among us who knows It truly, knows (what is meant by) "I know" and also what is meant by "I know It not."  This appears to be contradictory, but it is not.  In the previous chapter we learned that Brahman is "distinct from the known" and "beyond the unknown." The disciple, realizing this, says: "So far as mortal conception is concerned, I do not think I know, because I understand that It is beyond mind and speech; yet from the higher point of view, I cannot say that I do not know; for the very fact that I exist, that I can seek It, shows that I know; for It is the source of my being. }

Philosophically speaking, all the documents show differences with respect to the content they present. The first document deals with the duties of a government to create a society with peace. The second one is about the existence of God and about the belief in God. The third document deals with interactions among individuals and describes virtue. The last fragment is an abstract discussion about what knowledge is and can be classified as an epistemological text.

Despite the fact that we obtain similarities with respect to the Euclidean and Manhattan measure, we see that regarding the content of these documents, there is not a large overlapping. This prompts us to claim that these two methods are not focused on the meaning of the documents. Also, using the Jaccard and Cosine approaches, we obtain high similarities which leads us to the hypothesis that these measures also do not focus on the semantics. Thus, we claim that the similarities are mostly based on the syntactical structures of the documents.

Regarding the classification,  our findings reveal that the Multinomial Naive Bayes model yields the best prediction performance depicted in figures \ref{fig:oi6io} and \ref{fig:oii1o}. The Quran and Bible have the largest number of chapters in the corpus, and the Multinomial Naive Bayes is able to predict most of them with a very high accuracy, which is around 0.8584.  Random Forest (RF) follows with 0.737 of accuracy. K-nearest-neighbors (KNN), and Super Vector Machine (SVM) fails to distinguish the majority of chapters from the Asian sacred books. 

\textbf{Future research:} As future work, the plan is  to extend the set of our corpus by taking into consideration many other religions around the world, and therefore, more books. It could be for instance:
\begin{itemize}
	\item Inclusion of more books from Asian and Christian religions;
	\item Inclusion of the African tradition Scriptures and also the consideration of religions that are now obsolete. It will be very interesting to go deep into this study by exploring all the different properties that the distance measures use to analyze the meaning or semantic aspect of the sacred scriptures. 
\end{itemize}
In addition to that, since the focus of this study has been more on the lexical analysis, a further approach could include semantic. Given, for instance, the intrinsic dependence between words in raw texts, a much more informative feature engineering could be done before hand. That could include deriving from the bag of words, group of words/tokens as features instead of single tokens. Subsequently, deep learning approaches including Long Short-Term Memory (LSTM), and Convolution Recurrent Neural Network (CRNN) could be used to fit the data and extract, for instance, the contexts inside chapters of the books, categorize them, and run a contextual comparison analysis between (chapter of) different books. These models could also be used after effective training for completion of some of the sentences in the actual religious books.

\bibliographystyle{cas-model2-names}  
\bibliography{references}

\end{document}